\documentclass[]{eptcs}
\providecommand{\event}{GandALF 2011} 
\usepackage{mymath,proof,latexsym}

\begin{document}
\sloppy 
\hbadness=10000

\baselineskip=0.99\baselineskip

\pagestyle{myheadings}

\def\DB{\hbox{DB}}
\def\DBtwo{\hbox{DB2}}
\def\DBthree{\hbox{DB3}}
\def\DBfour{\hbox{DB4}}
\def\Nil{\hbox{nil}}
\def\Cons{\hbox{cons}}
\def\Null{\hbox{null}}
\def\Hd{\hbox{hd}}
\def\Tl{\hbox{tl}}
\def\Ifc{\hbox{ifc}}
\def\If{\hbox{if}\ }
\def\Then{\ \hbox{then}\ }
\def\Else{\ \hbox{else}\ }


\def\BRNI{{\hbox{BRNI}}}
\def\Rec{{\hbox{rec}}}
\def\FV{{\hbox{FV}}}
\def\FTV{{\hbox{FTV}}}
\def\Dom{{\hbox{Dom}}}
\def\Type{{\hbox{type}}}
\def\Bool{{\hbox{bool}}}
\def\Int{{\hbox{int}}}
\def\List{{\hbox{\ list}}}
\def\Pair{{\hbox{pair\ }}}
\def\Fst{{\hbox{fst\ }}}
\def\Snd{{\hbox{snd\ }}}
\def\Mgu{{\hbox{mgu}}}
\def\Fail{{\hbox{fail}}}
\def\If{{\hbox{if\ }}}
\def\Then{{\hbox{\ then\ }}}
\def\Else{{\hbox{\ else\ }}}
\def\BR{{\hbox{BR}}}
\def\BRlet{{\hbox{BR+let}}}
\def\Let{{\hbox{let\ }}}
\def\In{{\hbox{\ in\ }}}

\def\Tilde{\widetilde}

\title{Type Inference for Bimorphic Recursion}
\author{Makoto Tatsuta
\institute{National Institute of Informatics \\
2-1-2 Hitotsubashi, Tokyo 101-8430, Japan}
\email{tatsuta@nii.ac.jp}
\and 
Ferruccio Damiani
\institute{Dipartimento di Informatica, Universit\`a di Torino \\
Corso Svizzera 185, I-10149 Torino, Italy}
\email{damiani@di.unito.it}
}
\def\titlerunning{Type Inference for Bimorphic Recursion}
\def\authorrunning{M. Tatsuta \& F. Damiani}

\maketitle

\begin{abstract}
This paper proposes bimorphic recursion, which is restricted
polymorphic recursion such that every recursive call in the body of a
function definition has the same type.  Bimorphic recursion allows us
to assign two different types to a recursively defined function: one
is for its recursive calls and the other is for its calls outside its
definition.  Bimorphic recursion in this paper can be nested.  This
paper shows bimorphic recursion has principal types and decidable type
inference.  Hence bimorphic recursion gives us flexible typing for
recursion with decidable type inference.  This paper also shows that
its typability becomes undecidable because of nesting of recursions
when one removes the instantiation property from the bimorphic
recursion.
\end{abstract}

\def\emph{}

\section{Introduction}

The \emph{Hindley-Milner system}, which is called the \emph{ML type
system}~\cite{DamMil82} and the core of the type systems of
functional programming languages like SML, OCaml, and Haskell, is only
able to infer types for \emph{monomorphic recursion},
that is,
recursive function definitions 
where all the occurrences of recursive calls have the same simple type of the function definition. The problem of inferring types for
\emph{polymorphic recursion},
that is, recursive function definitions where different occurrences of
recursive calls have different
simple types that specialize the polymorphic type of the function
definition~\cite{Mee83,Myc84}, has been studied both by people
working on type systems~\cite{Henglein93,Kfoury93,Cop80,Jim96,Kfo+Per:LICS-1999,Ter+Aik:LICS-2006,Damiani:FI-07,Rittri95}
and by people working on abstract
interpretation~\cite{Monsuez92,Monsuez:SAS-1993,GoriLevi:VMCAI-2002,Gori03}.

Type inference for polymorphic recursion was shown to be undecidable
\cite{Henglein89,Kfoury93}.
For this reason, those programming languages do not
use polymorphic types for recursive definitions.
Haskell and OCaml allow polymorphic recursion only when we provide
type annotation.
On the other hand, 
a restricted form of polymorphic recursion could be useful for programming.
It is important theoretically as well as practically to
find some restriction such that
it is enough flexible and its type inference is decidable.

Henglein \cite{Henglein89} suggested that 
we have decidable type inference in
some restricted polymorphic recursion.
In this system,
only one recursive call in the body of the function
definition is allowed. In addition, recursive definitions must
not be nested. 
We call this recursion the single polymorphic recursion.
Type inference of the single polymorphic
recursion is reduced to semi-unification problems with a single
inequation, which are known to be decidable.

Our contribution is proposing bimorphic recursion and
proving that it has decidable type inference. 
Bimorphic recursion is an extension of the single polymorphic recursion.
In bimorphic recursion, each recursive call in the body of the function
definition must have the same type. Recursive definitions can be nested.
``Bimorphic'' means that a recursively defined function can have
two different types: one is for recursive calls in the body of the definition,
and another is for its calls of the function outside the body of the definition.
This paper shows that the system with bimorphic recursion has principal types 
and decidable type inference.

The idea for the type inference algorithm is based on an observation that
for a given recursive definition by bimorphic recursion,
we can infer its principal type by typing only the body of its definition.
We do not have to think of types of calls of the function outside the
body of its definition.
By this idea, our algorithm first handles the innermost recursive definition
and then goes to the second innermost recursive definition and so on.
According to the type inference algorithm in \cite{Henglein89,Henglein93},
our bimorphic recursion produces semi-unification problems, which are undecidable in general. 
Our idea enables us to reduce these semi-unification problems to
semi-unification problems with a single inequation, which are
decidable \cite{Henglein89}.
This algorithm can also work with polymorphic types in the let constructor.
The system with
bimorphic recursion and the polymorphic let constructor has
principal types and decidable type inference.

Because of nesting of recursions, this idea is so subtle that 
it becomes unavailable 
even by a small change of a typing system.
An example is the system with extended monomorphic recursions where
every recursive call has the same type that is an instantiation of
the type of the function definition.
Since every recursive call has the same type as the type of the function definition
in monomorphic recursion, the system is an extension of monomorphic recursion and
can type more expressions.
This system is also obtained from our bimorphic recursion type system
by removing
instantiation in the rule for recursion.
The system types less expressions than our bimorphic recursion type system.
Nonetheless
the same idea does not work for its type inference,
since the type of the function depends also on the types of its calls outside
the body of its definition.
Indeed this paper will show that its type inference is undecidable.
This will be proved by reducing semi-unification problems
to type inference.
This reduction is obtained
by refining the reduction of semi-unification problems to
the type inference of polymorphic recursion given in \cite{Henglein89}.

About ten years ago, building on results by Cousot~\cite{Cous:POPL-1997}, Gori and
Levi~\cite{GoriLevi:VMCAI-2002,Gori03} have developed a type
abstract interpreter that is able to type all the ML typable recursive
definitions and interesting examples of polymorphic recursion.
As pointed out in~\cite{CominiDamiani08}, the problem of
establishing whether Gori-Levi typability  is decidable is open.
Since our system is inspired by \cite{CominiDamiani08},
our bimorphic recursion type system can help to solve it.

Section 2 defines bimorphic recursion.
Section 3 gives its type inference algorithm and shows
bimorphic recursion has principal types and decidable type inference.
Section 4 discusses its extension to the polymorphic let
constructor. 
Section 5 studies bimorphic recursion without instantiation
and shows that its type inference is undecidable.
Section 6 concludes.

\section{The system $\BR$}

We will define the type system $\BR$ of bimorphic recursion.
We assume variables $x,y,z,\ldots$, and
constants $c,\ldots$. We have
expressions $e$ defined by 

$e ::= x | c | \lambda x.e | ee | \Rec\{x=e\}$.

These consist of $\lambda$-terms with constants and $\Rec$.
The expression $\Rec\{x=e\}$ means the recursively defined function $x$ 
by the equation $x=e$ where $e$ may contain recursive calls of $x$.
We will write $e[x:=e_1]$ for the expression obtained from $e$
by capture-avoiding substitution of $e_1$ for $x$.

We assume type variables $\alpha,\beta,\ldots$. We have
types $u,v,w$ defined by 

$u,v,w ::= \alpha | \Bool | \Int | u \imp u | u \times u | u \List $.

These consist of type variables, and the types of booleans, integers, functions,
cartesian products, and lists.

We will write $\FV(e)$ for the set of free variables in $e$ and
$\FTV(u)$ for the set of free type variables in $u$.

A substitution $s$ is defined as a function from type variables to types such that
$\{ \alpha | s(\alpha) \ne \alpha \}$ is finite.
We will write
$\Dom(s)=\{ \alpha | s(\alpha) \ne \alpha \}$.
We extend $s$ to types by defining $s(u)$ by 
s(\Bool)=\Bool, s(\Int)=\Int,
$s(u \imp v) = s(u) \imp s(v)$,
$s(u \times v) = s(u) \times s(v)$,
and $s(u \List)=s(u) \List$.
We will use $s,r$ for substitutions.
We will write $u[\alpha:=v]$ for the type obtained from $u$
by replacing $\alpha$ by $v$.

A type environment $U$ is the set $\{x_1:u_1,\ldots,x_n:u_n\}$ where 
$u_i = u_j$ for $x_i = x_j$.
We will write
$\FTV(U) = \FTV(u_1) \cup \ldots \cup \FTV(u_n)$ and
$\Dom(U) = \{ x_1,\ldots,x_n \}$.
We will use $U$ for type environments.

A judgment is of the form $U \prove e:u$. 
We will write $x_1:u_1,\ldots,x_n:u_n \prove e:u$
when $U$ is $\{x_1:u_1,\ldots,x_n:u_n\}$.
We will write $U,x:u \prove e:v$ for $U \cup \{x:u\} \prove e:v$.

We assume each constant $c$ is given its type denoted by
$\Type(c)$.

The system has the inference rules given in Figure \ref{fig:1}.
\begin{figure*}[t]
{\prooflineskip
\[
\infer[(var)]{U,x:u \prove x:u}{}
\qquad
\infer[(con)]{U \prove c:s(u)}{}
\quad (\Type(c)=u)
\\
\infer[(\imp I)]{U \prove \lambda x.e:u \imp v}{U,x:u \prove e:v}
\qquad
\infer[(\imp E)]{U \prove e_1e_2:u}{
	U \prove e_1:v \imp u
	&
	U \prove e_2:v
}
\\
\infer[(rec)]{U \prove \Rec\{x=e\}:s_2(u)}{
	U,x:s_1(u) \prove e:u
}
\quad (\Dom(s_1),\Dom(s_2) \subseteq \FTV(u)-\FTV(U))
\]
}
\caption{System $\BR$}\label{fig:1}
\end{figure*}

These rules form an extension of the simply typed $\lambda$-calculus
with the rules $(con)$ and $(rec)$.
By the rule $(con)$, the constant $c$ has the type $s(u)$
which is an instantiation $s(u)$ of the given type $u$.
By the rule $(rec)$, 
for the recursively defined function $x$ with its definition $x=e$,
we can use some general type $u$ to type the function.
First we have to show the body $e$ of its definition has
this type $u$ by assuming each recursive call of $x$ in $e$ has
the unique type $s_1(u)$ which is obtained from $u$ by instantiation with
a substitution $s_1$. Then we can say the defined function $x$
has the type $s_2(u)$ which is another instantiation of the type $u$.
The side condition guarantees that $s_1$ and $s_2$ change
only type variables that do not occur in $U$.

An expression $e$ is defined to be typable if $\prove e:u$ is provable for some type
$u$.
A type $u$ is defined to be a principal type for a term $e$ when
(1) $\prove e:u$ is provable, and (2)
if $\prove e:u'$ is provable, then there is some substitution $s$ such that
$s(u)=u'$.

\begin{Th}\label{th:principal}
There is a type inference algorithm for the type system $\BR$. That is,
there is an algorithm such that for a given term it
returns its principal type if the term is typable, and
it fails if the term is not typable.
\end{Th}

We will prove this theorem in the next section.
This theorem can be extended to a system with polymorphic let in Section 4.

We will write $u \to v \to w$ for $u \to (v \to w)$.
We assume the constants 
$\hbox{pair}$, $\hbox{fst}$, $\hbox{snd}$,
$\Nil$, $\Cons$, $\Hd$, $\Tl$, $\Null$, $0$,$1$,$2$, and $\Ifc$
with 
$\Type(\hbox{pair}) = \alpha_1 \to
\alpha_2 \to (\alpha_1 \times \alpha_2)$, $\Type(\hbox{fst})= (\alpha_1 \times
\alpha_2) \to \alpha_1$, $\Type(\hbox{snd}) = (\alpha_1 \times \alpha_2) \to
\alpha_2$,
$\Type(\Nil) = \alpha\List$,
$\Type(\Cons) = \alpha \to \alpha\List \to \alpha\List$,
$\Type(\Hd) = \alpha\List \to \alpha$,
$\Type(\Tl) = \alpha\List \to \alpha\List$,
$\Type(\Null) = \alpha\List \to \Bool$,
$\Type (0) = \Type(1) = \Type(2) = \Int$,
and $\Type(\Ifc) = \Bool \to \alpha \to \alpha \to \alpha$.
They are the pair, the first projection, the second
projection, the empty list,
the list construction, the head function for lists, 
the tail function for lists, 
the null function for the empty list,
three integers,
and the if-then-else statement respectively. We will use the following abbreviations.
\[
[e_1 . e_2] = \Cons\ e_1 e_2, \\ \relax
[\;] = \Nil, \\ \relax
[e_1, e_2, \ldots, e_n] = [e_1 . [e_2 . \ldots [e_n.[\;]] \ldots]], \\
\If e_1 \Then e_2 \Else e_3 = \Ifc\ e_1 e_2 e_3, \\
\]

We will explain bimorphic recursion by examples.

\begin{Eg}\label{eg:1}\rm
This example is a list doubling function
by using a dispatcher.
\[
\DB = \lambda x. (\DBtwo\ (\lambda y.y)) x, \\
\DBtwo = \Rec\{f_2=\lambda zw.\If (\Null\ w) \Then z [\;] \\
  \qquad \Else f_2 (\lambda xy.z(yx)) (\Tl\ w) (\lambda x.
[(\Hd\ w) . [(\Hd\ w) . x]])\}.
\]
According to informal meaning,
we have $\DB\ [0, 1, 2] = [0, 0, 1, 1, 2, 2]$ and
$(\DBtwo\ d) l = d(\DB\ l)$ where
$d$ is a dispatcher that takes the value $x$ and
several continuations $f_1,\ldots,f_n$ as its arguments and
returns $f_n(\ldots (f_1 x)\ldots )$.
The exact meaning is given by the following Haskell program.
\begin{verbatim}
  db x = (db2 (\y -> y)) x
  db2 :: ([b] -> a) -> [b] -> a
  db2 z w = if (null w) then z []
      else db2 (\x y -> z (y x)) (tail w) (\x -> (head w):(head w):x)
\end{verbatim}

The term $\DBtwo$ is not typable in ML since monomorphic recursion is not sufficient
for typing it.
This is typable in $\BR$ in the following way.
Let
\[
e_2 = \lambda zw.\If (\Null\ w) \Then z [\;] \Else f_2 (\lambda xy.z(yx)) (\Tl\ w) (\lambda x.[(\Hd\ w) . [(\Hd\ w) . x]]).
\]
We have
\[
f_2:(\beta\List \to (\beta\List \to \beta\List) \to \alpha) \to
\beta\List \to (\beta\List \to \beta\List) \to \alpha \\ \qquad
\prove 
e_2:(\beta\List \to \alpha) \to \beta\List \to \alpha.
\]
Hence by $(rec)$ with $s_1(\alpha)=(\beta\List \to \beta\List) \to \alpha$
and $s_2(\alpha)=\beta\List$,
we have
\[
\prove \DBtwo:(\beta\List \to \beta\List) \to \beta\List \to \beta\List
\]
and then we also have
\[
\prove \DB:\beta\List \to \beta\List.
\]
\end{Eg}

\begin{Eg}\label{eg:2}\rm
The following is an example for nesting of
bimorphic recursions.
This is obtained from $\DBtwo$ in the previous example by adding
some constant dummy task and writing them
by mutual recursion.
Let
\[
e_0 = [0, 1, 2], \\
e_3= \lambda zw.\If (\Null\ w) \Then (\lambda x.z [\;]) (f_4 (\lambda x.x) e_0) \\ \qquad
	\Else f_3 (\lambda xy.z(yx)) (\Tl\ w) (\lambda x.[(\Hd\ w) . [(\Hd\ w) . x]]).
\]
We want to define functions by the following mutual recursion.
\[
\DBthree = e_3[f_3:=\DBthree, f_4:=\DBfour], \\
\DBfour = \DBthree.
\]
The functions $\DBthree$ and $\DBfour$ behave in the same way
as $\DBtwo$ except
that the additional task $f_4 (\lambda x.x) e_0$ calculates the doubled list of the fixed list $e_0$,
and its resulting value is thrown away.
We can actually define these functions by using nests of bimorphic recursions
as follows:
\[
\DBfour = \Rec\{f_4=\Rec\{f_3=e_3\}\}, \\
\DBthree = \Rec\{f_3=e_3[f_4:=\DBfour]\}.
\]
Since the body of each recursion has only one recursive call, they are
bimorphic recursion. This kind of patterns 
cannot simulate full polymorphic recursion 
because of the variable side condition.

Note that the following $\DBthree'$ does not work,
since the recursive call $f_3$ occurs twice with different types in the body and
it is not bimorphic recursion.
\[
\DBthree' = \Rec\{f_3=e_3[f_4:=f_3]\}.
\]
We also note that the following $\DBthree''$ does not work,
since the recursive call $f_3$ occurs twice with different types in the body and
it is not bimorphic recursion.
\[
\DBthree'' = \Rec\{f_3=e_3[f_4:=\Rec\{f_4=f_3\}]\}.
\]

We can type $\DBthree$ and $\DBfour$ in our system in the following way.
We have
\[
f_4:(\Int\List \to \Int\List) \to \Int\List \to \Int\List, \\ 
f_3:(\beta\List \to (\beta\List \to \beta\List) \to \alpha) \to
  \beta\List \to (\beta\List \to \beta\List) \to \alpha \\ \qquad
  \prove 
  e_3:(\beta\List \to \alpha) \to \beta\List \to \alpha.
\]
By $(rec)$ rule, we have
\[
f_4:(\Int\List \to \Int\List) \to \Int\List \to \Int\List \\ \qquad \prove 
  \Rec\{f_3=e_3\}:
  (\beta\List \to \beta\List) \to (\beta\List \to \beta\List).
\]
By $(rec)$, we have
\[
\prove \DBfour:(\Int\List \to \Int\List) \to \Int\List \to \Int\List.
\]
We also have
\[
f_3:(\beta\List \to (\beta\List \to \beta\List) \to \alpha) \to
  \beta\List \to (\beta\List \to \beta\List) \to \alpha \\ \qquad
  \prove 
  e_3[f_4:=\DBfour]:(\beta\List \to \alpha) \to \beta\List \to \alpha
\]
and by $(rec)$ we have
\[
\prove \DBthree:(\beta\List \to \beta\List) \to \beta\List \to \beta\List.
\]
\end{Eg}

\section{Type Inference Algorithm}

This section gives our type inference algorithm for bimorphic recursion,
and proves its correctness.

A principal typing is defined as
a judgment $U \prove e:u$ 
when
(1) $U \prove e:u$ is provable, and (2)
if $U' \prove e:u'$ is provable, then there is some substitution $s$ such that
$s(U)=U'$ and $s(u)=u'$.

Given types $u,v$, we write $u=v$ to mean $u$ is $v$.
We write $u \le v$ to mean there is some substitution $s$ such that
$s(u)=v$.
A unification problem
is defined as 
a set of equations of the form $u=v$.
We say a substitution $s$ is a unifier of the unification problem
$\{ u_1=u_1', \ldots, u_n=u_n' \}$
when $s(u_1)=s(u_1'), \ldots, s(u_n)=s(u_n')$ hold.
A semi-unification problem
is defined as
a set of equations of the forms $u=v$ and inequations of the form $u \le v$.
We say a substitution $s$ is a semiunifier of the semiunification problem
$\{ u_1=u_1', \ldots, u_n=u_n', v_1 \le v_1', \ldots, v_m \le v_m' \}$
when $s(u_1)=s(u_1'), \ldots, s(u_n)=s(u_n'), s(v_1) \le s(v_1'), \ldots, s(v_m) \le s(v_m')$ hold.
A typing problem
is defined as
the judgment $x_1:u_1,\ldots,x_n:u_n \prove e:u$.
We say a substitution $s$ is a solution of the typing problem
$x_1:u_1,\ldots,x_n:u_n \prove e:u$
when $x_1:s(u_1),\ldots,x_n:s(u_n) \prove e:s(u)$ is provable.

We will write $s(U)$ for $\{x_1:s(u_1),\ldots,x_n:s(u_n)\}$
when $U$ is $\{x_1:u_1,\ldots,x_n:u_n\}$.
We will use vector notation $\vec u$ for a sequence $u_1,\ldots,u_n$.
We will sometimes denote the set $\{ u_1,\ldots,u_n \}$ by $\vec u$.

When we take Henglein's algorithm \cite{Henglein89,Henglein93} for our system,
we have the following algorithm $E'$ that produces a semi-unification
problem from a given judgment such that
$E'(U \prove e:u)=E_0$ and $s$ is a most general unifier of $E_0$ if
and only if $s(U) \prove e:s(u)$ is a principal typing.
\[
E'(U,x:u \prove x:v) = \{ u=v \},\\
E'(U \prove x:u) = \{ \Bool=\Int \} \hbox{\ where\ } x \not\in \Dom(U),\\
E'(U \prove c:u) = \{ u=v \} \hbox{\ where\ } \Type(c)=v, \\
E'(U \prove \lambda x.e_1:u) = 
	E'(U,x:\alpha \prove e_1:\beta) \cup \{ \alpha \imp \beta=u \} 
	\hbox{\ where\ }\\
\qquad \alpha,\beta \hbox{\ fresh type variables},\\
E'(U \prove e_1e_2:u) = E'(U \prove e_1:\alpha \imp u) \cup E'(U \prove e_2:\alpha) \hbox{\ where\ } \\
\qquad \alpha \hbox{\ a fresh type variable},\\
E'(U \prove \Rec\{x=e_1\}:u)=E_1 
	\cup \{\alpha \times \vec u \le \beta \times \vec u, 
	\alpha \times \vec u \le u \times \vec u \} \\
\qquad \hbox{where\ } \alpha,\beta \hbox{\ fresh type variables},\\
\qquad U=\{ x_1:u_1,\ldots, x_n:u_n \}, \\
\qquad \vec u=u_1 \times \ldots \times u_n, \\
\qquad E'(U,x:\beta \prove e_1:\alpha)=E_1.
\]
Note that the type inference for the rec construct produces inequations.

This does not give a decidable type inference for our system, since
it may produce a semi-unification problem with two or more inequations.
For example, when we apply it to the function $\DBfour$ in Example \ref{eg:2},
we have four inequations in $E'(\prove \DBfour:\alpha)$.
By some property of the semi-unification problem,
we can eliminate two inequations.
Then the semi-unification problem becomes
\[
\{ \alpha_1 \le \beta_1, \alpha_2 \times \beta_1 \le \beta_2 \times \beta_1 \}
\cup E_0
\]
where $E_0$ is some set of equations.
In general, semi-unification problems with two inequations are undecidable \cite{Henglein89}.
Our key idea is that for bimorphic recursion, 
we can always divide the semi-unification problem
$
\{ \alpha_1 \le \beta_1, \alpha_2 \times \beta_1 \le \beta_2 \times \beta_1 \}
\cup E_0$
into two problems
$\{ \alpha_1 \le \beta_1 \} \cup E_1$
and
$\{ \alpha_2 \times \beta_1 \le \beta_2 \times \beta_1 \} \cup E_2$
where $E_1$ and $E_2$ are some sets of equations 
such that
in order to solve
$
\{ \alpha_1 \le \beta_1, \alpha_2 \times \beta_1 \le \beta_2 \times \beta_1 \}
 \cup E_0$
it is sufficient to first solve
$\{ \alpha_2 \times \beta_1 \le \beta_2 \times \beta_1 \} \cup E_2$
and
then solve
$\{ \alpha_1 \le \beta_1 \} \cup E_1$.
This idea reduces those  semi-unification problems into
semi-unification problems with a single inequation and
gives an algorithm of solving them, since
semi-unification problems with a single inequation are decidable \cite{Henglein89}.

This idea is based on the observation that
bimorphic recursion can be typed locally.
We explain this observation.
We first tried to find an algorithm like $E'$ which behaves as follows:
in the same way 
as type inference algorithms for simply typed lambda calculus,
when a term $e$, a type $u$, and a type environment $U$ are given, 
the algorithm chases the proof of
$U \prove e:u$ upward from the conclusion,
and produces a set of equations between types such that 
the existence of its unifier $s$ is equivalent to
the provability of $s(U) \prove e:s(u)$.
Then we had difficulty for the $(rec)$ rule.

The idea is that we follow the above algorithm but we handle
the $(rec)$ rule in a separate way.
First we choose an uppermost $(rec)$ rule in the proof:
\[
\infer*[\pi_2]{}{
\infer[(rec)]{U \prove \Rec\{x=e\}:s_2(u)}{
\infer*[\pi_1]{U,x:s_1(u) \prove e:u}{}
}}
\]
Let $V$ be $\FTV(u)-\FTV(U)$. 
We will use $\pi_3$ to denote the subproof with the $(rec)$ rule and $\pi_1$.
The subproof $\pi_2$ cannot access $V$ because $s_2$ hides $V$ and
$U$ does not have any information of $V$.
Hence type inference for $\pi_3$ can be done separately from $\pi_2$.
Since $\pi_3$ has only one $(rec)$ rule, 
the type inference for $\pi_3$ is reduced to
a semi-unification problem with a single inequation.
Hence
type inference for $\pi_3$ is possible since there is an algorithm
solving a semi-unification problem with a single inequation.
By this, we will have a most general semiunifier $s$ and a principal 
type $v$ of $\Rec\{x=e\}$.
Then our type inference is reduced to type inference of the proof $\pi_4$:

\[
\infer*[s(\pi_2)]{}{
\infer[(axiom)]{s(U) \prove \Rec\{x=e\}:s_1(v)}{}
}
\]
for some $s_1$
where $s(\pi_2)$ denotes the proof obtained from $\pi_2$ by replacing
every judgment $U_1 \prove e_1:u_1$ by $s(U_1) \prove e_1:s(u_1)$
and the rule $(axiom)$ denotes a temporary axiom.
Since this reduction eliminates one $(rec)$ rule, by repeating this
reduction, we can reduce our type inference problem to type inference
problem for some term without the $(rec)$ rule. Hence we can complete type inference by solving it 
with the type inference algorithm for the simply typed lambda calculus.

We will write $\FTV(a_1,\ldots,a_n)$ for $\FTV(a_1) \cup \ldots \cup \FTV(a_n)$
when $a_i$ is a type or a type environment.
We define $\FTV(\{ u_1=u_1', \ldots, u_n=u_n' \})$ as
$\FTV(u_1,u_1',\ldots,u_n,u_n')$.

For a substitution $s$, a type variable $\alpha$, and 
a type $u$, the substitution $s[\alpha:=u]$ is defined by
$s[\alpha:=u](\alpha)=u$, and
$s[\alpha:=u](\beta)=s(\beta)$ if $\beta \ne \alpha$.

For a substitution $s$ and a set $V$ of type variables,
the substitution $s|_V$ is defined by
$s|_V(\alpha) = s(\alpha)$ if $\alpha \in V$ and
$s|_V(\alpha) = \alpha$  if $\alpha \not\in V$.

For substitutions $s_1,s_2$, the substitution $s_1s_2$ is defined
by $s_1s_2(\alpha)=s_1(s_2(\alpha))$.

For substitutions $s_1,s_2$ and a set $V$ of type variables,
$s_1 =_V s_2$ is defined to hold if $s_1(\alpha)=s_2(\alpha)$ for $\alpha \in V$. We will say $s_1 =s_2$ for $V$ when $s_1 =_V s_2$.

We will write $1$ for the identity substitution, that is, $1(\alpha)=\alpha$.

For a semiunification problem $S$, we say $s$ is a most general 
semiunifier of $S$ when $s$ is a semiunifier of $S$ and
for every semiunifier $r$ of $S$ there is a substitution $r'$ such that
$r's=r$.

We will write $\Mgu$ for the algorithm that returns a most general semiunifier
for a semiunification problem with a single inequation,
that is,
$\Mgu( \{ u_1=u_1', \ldots, u_n=u_n', v \le v' \} ) = s$ if
$s$ is a most general semiunifier of the semiunification problem
$\{ u_1=u_1', \ldots, u_n=u_n', v \le v' \}$, and
$\Mgu( \{ u_1=u_1', \ldots, u_n=u_n', v \le v' \} ) = \Fail$ if
no semiunifier exists for
the semiunification problem
$\{ u_1=u_1', \ldots, u_n=u_n', v \le v' \}$.
We assume $\Mgu$ uses fresh variables.

\begin{Def}[Type Inference Algorithm]\rm
In Figure \ref{fig:3},
we define an algorithm $E$ that takes a typing problem as its input and
returns a pair of a unification problem and a substitution as its outputs.
That is, $E(U \prove e:u)=(E_0, s_0)$ where $E_0$ is a
unification problem.
The algorithm $E$ assumes fresh variables.
Fresh variables are maintained globally and
$E(U \prove e:u)$ may return an answer with different fresh variables
depending on its global context.
\begin{figure*}[t]
\[
E(U,x:u \prove x:v) = (\{ u=v \}, 1),\\
E(U \prove x:u) = ( \{ \Bool=\Int \}, 1) \hbox{\ where\ } x \not\in \Dom(U),\\
E(U \prove c:u) = ( \{ u=v \}, 1) \hbox{\ where\ } \Type(c)=v, \\
E(U \prove \lambda x.e_1:u) = (E_1 \cup \{ s_1(\alpha \imp \beta)=s_1(u) \}, 
	s_1) \hbox{\ where\ }\\
\qquad \alpha,\beta \hbox{\ fresh type variables},\\
\qquad E(U,x:\alpha \prove e_1:\beta)=(E_1,s_1), \\
E(U \prove e_1e_2:u) = (s_2(E_1) \cup E_2, s_2s_1) \hbox{\ where\ } \\
\qquad \alpha \hbox{\ a fresh type variable},\\
\qquad E(U \prove e_1:\alpha \imp u)= (E_1,s_1), \\
\qquad E(s_1(U) \prove e_2:s_1(\alpha))=(E_2,s_2), \\
E(U \prove \Rec\{x=e_1\}:u)=(\{ s_2s_1(u)=s_2s_1(\alpha) \}, s_2s_1) \hbox{\ where\ } \\
\qquad \alpha,\beta \hbox{\ fresh type variables},\\
\qquad E(U,x:\beta \prove e_1:\alpha)=(E_1,s_1), \\
\qquad U=\{ x_1:u_1,\ldots, x_n:u_n \}, \\
\qquad \vec u=u_1 \times \ldots \times u_n, \\
\qquad s_2=\Mgu(E_1 \cup \{ s_1(\alpha \times \vec u) \le s_1(\beta \times \vec u) \}), \\
E(U \prove \Rec\{x=e_1\}:u)=(\{ \Bool=\Int \}, 1) \hbox{\ where\ } \\
\qquad \alpha,\beta \hbox{\ fresh type variables},\\
\qquad E(U,x:\beta \prove e_1:\alpha)=(E_1,s_1), \\
\qquad U=\{ x_1:u_1,\ldots, x_n:u_n \}, \\
\qquad \vec u=u_1 \times \ldots \times u_n, \\
\qquad \Mgu(E_1 \cup \{ s_1(\alpha \times \vec u) \le s_1(\beta \times \vec u) \})=\Fail.
\]
\caption{Type Inference Algorithm}\label{fig:3}
\end{figure*}
\end{Def}

When $E(U \prove e:u)=(E_0,s_0)$, the typing problem $U \prove e:u$
is reduced to the unification problem $E_0$. 
The substitution $s_0$ gives a partial solution
of the typing problem, that is, with a unifier $s$ of $E_0$,
the typing problem $U \prove e:u$ has a solution $ss_0$.

\begin{Prop}\label{prop:instantiate}
If $U \prove e:u$ is provable, then $s(U) \prove e:s(u)$ is provable
for any $s$.
\end{Prop}

{\em Proof.}
By induction on the proof. We consider cases according to the last rule.

Case $(var)$. Suppose $U,x:u \prove x:u$. We have $s(U),x:s(u) \prove x:s(u)$.

Case $(con)$. Suppose $U \prove c:s_1(u)$. We have $s(U) \prove c:ss_1(u)$.

Case $(\imp I)$. Suppose
\[
\infer{U \prove \lambda x.e:u \imp v}{
\infer*{U,x:u \prove e:v}{}
}
\]
By IH, we have $s(U),x:s(u) \prove e:s(v)$. Hence we have the claim.

Case $(\imp E)$ is similar to Case $(\imp I)$.

Case $(rec)$. Suppose
\[
\infer{U \prove \Rec\{x=e\}:s_2(u)}{
\infer*{U,x:s_1(u) \prove e:u}{}
}
\]
Let $\vec\alpha$ be $\FTV(u)-\FTV(U)$.
We can assume $\vec\alpha$ are fresh. Hence
we have $\Dom(s) \cap \vec\alpha =
s(U) \cap \vec\alpha = \emptyset$.

Let $s_1'= (ss_1) |_{\vec\alpha}$ and $s_2'= (ss_2) |_{\vec\alpha}$.
We have $ss_1(u)=s_1's(u)$ since
$ss_1(\alpha_i)=s_1'(\alpha_i)=s_1's(\alpha_i)$ and
$ss_1(\beta)=s(\beta)=s_1's(\beta)$ for $\beta \in \FTV(U)$.
Similarly we have $ss_2(u)=s_2's(u)$.

By IH, we have $s(U),x:ss_1(u) \prove e:s(u)$.
Hence we have $s(U),x:s_1's(u) \prove e:s(u)$.
We have $\Dom(s_1') \subseteq \FTV(s(u))-\FTV(s(U))$ since
$\Dom(s_1') \subseteq \vec\alpha$ holds and
$\alpha_i \in \FTV(s(u))$ and $\alpha_i \not\in \FTV(s(U))$ show
$\vec\alpha \subseteq \FTV(s(u))-\FTV(s(U))$.
Similarly we have $\Dom(s_2') \subseteq \FTV(s(u))-\FTV(s(U))$.
By the rule $(rec)$, we have $s(U) \prove \Rec\{x=e\}:s_2's(u)$.
Hence we have the claim.
$\Box$

We define
$s(\{ \alpha_1,\ldots,\alpha_n \})$ as $\{ s(\alpha_1),\ldots,s(\alpha_n) \}$.

\begin{Th}\label{th:1}
If 
the typing problem $U \prove e:u$ has a solution $s$,
$V$ is a finite set of type variables, $\FTV(U) \cup \FTV(u) \subseteq V$,
and $E(U \prove e:u)=(E_0,s_0)$,
then there is a unifier $s_0'$ of $E_0$ such that $s_0's_0 =_V s$.
\end{Th}

{\em Proof.}
By induction on $e$. We consider cases according to $e$.

Case $e=x$. We can suppose $U$ is $U_1,x:v$. We have $s(v)=s(u)$ and
$(E_0,s_0)=(\{ v=u \}, 1)$. We can take $s_0'=s$.

Case $e=c$. Let $\Type(c)=v$. 
We have $s_1(v)=s(u)$ for some $s_1$.
We can assume $\FTV(v)$ is fresh.
Then we have $\FTV(v) \cap V = \emptyset$.
We have $(E_0,s_0)=(\{ u=v \}, 1)$.
We can define $s_0'$ by $s_0'(\alpha)=s(\alpha)$ for $\alpha \in V$
and $s_0'(\alpha)=s_1(\alpha)$ for $\alpha \in \FTV(v)$.
$s_0'$ is a unifier of $E_0$ since $s_0'(v)=s_1(v)$ and $s_0'(u)=s(u)$.
$s_0's_0=_V s$ since $s_0'=_V s$.

Case $e=\lambda x.e_1$.
We suppose $s(U),x:v_1 \prove e_1:v_2$ and $v_1 \imp v_2 = s(u)$.
Let 
$\Tilde s=s[\alpha:=v_1,\beta:=v_2]$.
The typing problem $U,x:\alpha \prove e_1:\beta$ has a solution
$\Tilde s$.
Let $V_1=V\cup \{\alpha,\beta\}$.
By induction hypothesis for $e_1$ with $V_1$,
there is a unifier $s_1'$ of $E_1$ such that $s_1's_1=_{V_1} \Tilde s$.
We can take $s_0'=s_1'$.
$s_0'$ is a unifier of $E_0$ since
$s_1'$ is a unifier of $E_1$, and
$s_1's_1=\Tilde s$ for $\alpha,\beta,u$.
$s_0's_0=_V s$ since $s_1's_1=\Tilde s=s$ for $V$.

Case $e=e_1e_2$.
We suppose $s(U) \prove e_1:v \imp s(u)$ and $s(U) \prove e_2:v$.
Let $\Tilde s = s[\alpha:=v]$.
The typing problem $U \prove e_1:\alpha \imp u$ has a solution $\Tilde s$.
Let $V_1=V\cup \{ \alpha \}$.
By induction hypothesis for $e_1$ with $V_1$, 
we have a unifier $s_1'$ of $E_1$ such that
$s_1's_1 =_{V_1} \Tilde s$.
We have $\Tilde s(U) \prove e_2:\Tilde s(\alpha)$.
Hence the typing problem $s_1(U) \prove e_2:s_1(\alpha)$ has
a solution $s_1'$.
Let $V_2 = s_1(V_1) \cup \FTV(E_1)$.
By induction hypothesis for $e_2$ with $V_2$,
$E_2$ has a unifier $s_2'$ and $s_2's_2=_{V_2}s_1'$.
We can take $s_0'=s_2'$.
$s_0'$ is a unifier of $E_0$ since
$s_2'$ is a unifier of $E_2$, and
$s_2's_2$ is a unifier of $E_1$ by
$s_2's_2=s_1'$ for $\FTV(E_1)$.
$s_0's_0=_V s$ since
$s = \Tilde s = s_1's_1 = s_2's_2s_1$ for $V$.

Case $e=\Rec\{x=e_1\}$.
We suppose $s(U),x:r_1(v) \prove e_1:v$ and
$s(u)= r_2(v)$ where $\Dom(r_1), \Dom(r_2) \subseteq \FTV(v)-\FTV(s(U))$.
We can assume $\FTV(v)-\FTV(s(U))$ is fresh and 
$(\FTV(v)-\FTV(s(U))) \cap \FTV(s(V)) = \emptyset$.
Hence we have $r_1s=_V s$ and $r_2s=_V s$.
Let $\Tilde s=s[\alpha:=v, \beta:=r_1(v)]$.
We have $\Tilde s(U),x:\Tilde s(\beta) \prove e:\Tilde s(\alpha)$.
Hence the typing problem $U,x:\beta \prove e:\alpha$ has a solution $\Tilde s$.
Let $V_1$ be $V\cup \{\alpha,\beta\}$.
By induction hypothesis for $e_1$ with $V_1$, 
we have a unifier $s_1'$ of $E_1$ such that $s_1's_1=_{V_1}\Tilde s$.
Then $s_1'$ is a semiunifier of 
$E_1 \cup \{ s_1(\alpha \times \vec u) \le s_1(\beta \times \vec u) \}$,
since
$r_1s_1's_1(\alpha \times \vec u) = s_1's_1(\beta \times \vec u)$, which
is proved as follows:
$r_1s_1's_1(\alpha)=s_1's_1(\beta)$ since 
$s_1's_1 = \Tilde s$ for $\alpha,\beta$.
$r_1s_1's_1(\vec u)=s_1's_1(\vec u)$ since 
$s_1's_1(\vec u) = \Tilde s(\vec u) = s(\vec u)$ and 
$r_1s(\vec u) = s(\vec u)$.

Since $s_2$ is a most general semiunifier, we have $s_2'$ such that
$s_2's_2=s_1'$.
We can take $s_0'=r_2s_2'$.
$s_0's_2s_1(u)=s_0's_2s_1(\alpha)$ since
$r_2s_2's_2s_1 = r_2s_1's_1 = r_2\Tilde s$ for $\alpha$ and $\FTV(u)$,
$r_2\Tilde s(u)=r_2s(u)=s(u)$, and $r_2\Tilde s(\alpha)= r_2(v)=s(u)$.
$s_0's_0 =_V s$ since
$r_2s_2's_2s_1=r_2s_1's_1=_V r_2\Tilde s=_V r_2s=_V s$.
$\Box$

\begin{Th}\label{th:2}
If $E(U \prove e:u)=(E_0,s_0)$ and $s$ is a unifier of $E_0$,
then
$ss_0$ is a solution of the typing problem $U \prove e:u$.
\end{Th}

{\em Proof.}
By induction on $e$. We consider cases according to $e$.

Case $e=x$. 
Since $\{ \Bool=\Int \}$ does not have any unifier, we have $x \in \Dom(U)$.
We suppose $U$ is $U_1,x:u$. 
We have $(E_0,s_0)=(\{ u=v \},1)$ and $s(u)=s(v)$.
By the rule $(var)$, we have $s(U_1),x:s(u) \prove x:s(v)$ and $ss_0$ is a solution.

Case $e=c$. We suppose $\Type(c)=v$.
We have $(E_0,s_0)=(\{ u=v \},1)$ and $s(u)=s(v)$.
By the rule $(con)$, we have $s(U) \prove c:s(u)$ and $ss_0$ is a solution.

Case $e=\lambda x.e_1$. Since $s$ is a unifier of $E_1$,
by induction hypothesis for $e_1$, 
$ss_1$ is a solution of the typing problem $U,x:\alpha \prove e_1:\beta$.
Hence $ss_1(U),x:ss_1(\alpha) \prove e_1:ss_1(\beta)$.
By the rule $(\imp I)$, 
we have $ss_1(U) \prove \lambda x.e_1:ss_1(\alpha \imp \beta)$.
Since $s$ is a unifier of $\{ s_1(\alpha \imp \beta)=s_1(u) \}$,
we have $ss_1(\alpha \imp \beta) = ss_1(u)$.
Hence we have $ss_1(U) \prove e:ss_1(u)$.
Therefore $ss_0$ is a solution of the typing problem $U \prove e:u$.

Case $e=e_1e_2$. Since $s$ is a unifier of $E_2$,
by induction hypothesis for $e_2$,
$ss_2$ is a solution of the typing problem $s_1(U) \prove e_2:s_1(\alpha)$.
Then $ss_2s_1(U) \prove e_2:ss_2s_1(\alpha)$.
Since $ss_2$ is a unifier of $E_1$,
by induction hypothesis for $e_1$,
$ss_2s_1$ is a solution of the typing problem $U \prove e_1:\alpha \imp u$.
Then $ss_2s_1(U) \prove e_1:ss_2s_1(\alpha) \imp ss_2s_1(u)$.
By the rule $(\imp E)$, we have $ss_2s_1(U) \prove e_1e_2:ss_2s_1(u)$.
Therefore $ss_0$ is a solution of the typing problem $U \prove e:u$.

Case $e=\Rec\{x=e_1\}$.
Since $\{ \Bool=\Int \}$ does not have any unifier, we have
$s_2=\Mgu(E_1 \cup \{ s_1(\alpha \times \vec u) \le s_1(\beta \times \vec u) \})$.
Since $s_2$ is a unifier of $E_1$, 
by induction hypothesis for $e_1$,
we have $s_2s_1(U),x:s_2s_1(\beta) \prove e_1:s_2s_1(\alpha)$.
Since $s_2s_1(\alpha \times \vec u) \le s_2s_1(\beta \times \vec u)$ holds,
we have $s_3$ such that
$s_3s_2s_1(\alpha \times \vec u) = s_2s_1(\beta \times \vec u)$.
We can suppose $\Dom(s_3) \subseteq \FTV(s_2s_1(\alpha))$.
Then we have $s_3s_2s_1(\alpha)=s_2s_1(\beta)$
and $s_3s_2s_1(\vec u)=s_2s_1(\vec u)$.
Hence we have $s_2s_1(U),x:s_3s_2s_1(\alpha) \prove e_1:s_2s_1(\alpha)$
and $\Dom(s_3) \subseteq \FTV(s_2s_1(\alpha))-\FTV(s_2s_1(U))$.
By the rule $(rec)$, we have $s_2s_1(U) \prove \Rec\{x=e_1\}:s_2s_1(\alpha)$.
By Proposition \ref{prop:instantiate},
we have $ss_2s_1(U) \prove e:ss_2s_1(\alpha)$.
Since $s$ is a unifier of $\{ s_2s_1(u)=s_2s_1(\alpha) \}$,
we have $ss_2s_1(u)=ss_2s_1(\alpha)$.
Then $ss_2s_1(U) \prove e:ss_2s_1(u)$.
Hence $ss_0$ is a solution of the typing problem $U \prove e$.
$\Box$

{\bf Proof of Theorem \ref{th:principal}.}
We define the algorithm as follows.
Suppose $e$ is given. We will provide its principal type if $e$ has
a type and return the fail if $e$ does not have any type.
Let $\alpha$ be a fresh type variable.
Let $E(\prove e:\alpha) = (E_0,s_0)$.
If $E_0$ does not have any unifier, we return the fail.
Otherwise let $s_1$ be a most general unifier of $E_0$.
Let $u$ be $s_1s_0(\alpha)$. We return $u$.

We will show that if the algorithm fails then $e$ does not has any type.
We assume the algorithm fails and $\prove e:v$. We will show a contradiction.
We define $r$ by $r(\alpha)=v$. Then $r$ is a solution of
the typing problem $\prove e:\alpha$.
By Theorem \ref{th:1} for $\prove e:\alpha$ and $r$,
we have a unifier $r'$ of $E_0$.
Hence the algorithm does not fail, which leads to a contradiction.

We will show that if the algorithm returns a type then it is
a principal type. Suppose the algorithm returns $u$.
We will show $u$ is a principal type of $e$.
First we will show $\prove e:u$.
By Theorem \ref{th:2} for $(E_0,s_0)$ and $s_1$, 
$s_1s_0$ is a solution of the typing problem
$\prove e:\alpha$. Hence $\prove e:s_1s_0(\alpha)$ and $\prove e:u$.
Next we will show $\prove e:v$ implies $u \le v$.
We define $r$ by $r(\alpha)=v$. Then $r$ is a solution of
the typing problem $\prove e:\alpha$.
By Theorem \ref{th:1} for $\prove e:\alpha$ and $r$ with $V=\{ \alpha \}$,
we have a unifier $r'$ of $E_0$ such that $r's_0=_V r$.
Since $s_1$ is a most general unifier of $E_0$, 
we have $s_2s_1=r'$ for some $s_2$.
We have $s_2(u)=v$ since $s_2(u)=s_2s_1s_0(\alpha)= r's_0(\alpha)=r(\alpha)=v$.
$\Box$

\section{Bimorphic Recursion and Polymorphic Let}

The system $\BR$ of bimorphic recursion can be extended with
the standard polymorphic let constructor.
The resulting system also has 
principal types and decidable type inference.
We will discuss this extension.

We will define the type system $\BRlet$.

The types in $\BR$ will be called mono types.
Mono types $u,v,w$ are defined
by

$u,v,w ::= \alpha | \Bool | \Int | u \imp u | u \times u | u \List $.

Type types in $\BRlet$ include polymorphic types.
Types $A,B,C$ are defined by

$A,B,C ::= u | \forall \alpha.A$.

A type environment $U$ is the set $\{x_1:A_1,\ldots,x_n:A_n\}$ where 
$A_i = A_j$ for $x_i = x_j$.

A judgment is of the form $U \prove e:u$.

A mono type substitution $s$ is 
a function from type variables to mono types such that
$\{ \alpha | s(\alpha) \ne \alpha \}$ is finite.

The inference rules are those in $\BR$ except that
the rule $(var)$ is replaced by the following $(var-P)$,
a mono type substitution is used instead of a substitution in
the rule $(rec)$,
and the following rule $(let)$ is added.

{\prooflineskip 
\[
\infer[(var-P)]{U,x:\forall \alpha_1\ldots\alpha_n.u \prove x:s(u)}{}
\quad (\Dom(s) \subseteq \{\alpha_1,\ldots,\alpha_n\})
\\
\infer[(let)]{U \prove \Let x=e_1 \In e_2:u}{
	U \prove e_1:v
	&
	U,x:\forall \alpha_1\ldots\alpha_n.v \prove e_2:u
}
(\alpha_1,\ldots,\alpha_n \in \FTV(v)-\FTV(U))
\]
}

\begin{Th}\label{th:principal2}
There is a type inference algorithm for the type system $\BRlet$. That is,
there is an algorithm such that for a given term it
returns its principal type if the term is typable, and
it returns the fail if the term is not typable.
\end{Th}

This is proved by extending
the type inference procedure $E$ for $\BR$ in Section 3 
to $\BRlet$ by
replacing the variable case by
\[
E(U,x:\forall\vec\alpha.u \prove x:v)=(\{u[\vec\alpha:=\vec\beta]=v\},1) \hbox{\ where\ } \\
\qquad \vec\beta \hbox{\ fresh type variables},
\]
and adding the following let cases:
\[
E(U \prove \Let x=e_1 \In e_2:u)=(E_2,s_3s_2s_1) \hbox{\ where\ } \\
\qquad \alpha \hbox{\ a fresh type variable},\\
\qquad E(U \prove e_1:\alpha)=(E_1,s_1),\\
\qquad \Mgu(E_1)=s_2, \\
\qquad \vec\beta=\FTV(s_2s_1(\alpha))-\FTV(s_2s_1(U)),\\
\qquad E(s_2s_1(U),x:\forall\vec\beta.s_2s_1(\alpha) \prove e_2:s_2s_1(u))=(E_2,s_3), \\
E(U \prove \Let x=e_1 \In e_2:u)=(\{ \Bool=\Int \}, 1) \hbox{\ where\ } \\
\qquad \alpha \hbox{\ a fresh type variable},\\
\qquad E(U \prove e_1:\alpha)=(E_1,s_1),\\
\qquad \Mgu(E_1)=\Fail.
\]

\section{Bimorphic Recursion with No Instantiation}

This section discusses the type system $\BRNI$
which is obtained from the type system $\BR$ by
removing the instantiation property.
We will show the type inference for $\BRNI$ is
undecidable by reducing semi-unification problems to it.

Semiunification terms $M,N$ are defined by $M,N ::= \alpha | M \times M$ where $\alpha$ is a type variable.
Note that a semiunification term is a type of $\BR$.

The following fact is well known for semi-unification problems.

\begin{Th}[\cite{Henglein89}]\label{th:semiunification}
The existence of a semiunifier of 
the set of two inequations is undecidable.
That is, there is no algorithm that decides if there is some $s$
such that $s_1(s(M_1))=s(N_1)$ and $s_2(s(M_2))=s(N_2)$ for some $s_1,s_2$
for a given semiunification problem $\{ M_1 \le N_1, M_2 \le N_2 \}$.
\end{Th}

We define the type system $\BRNI$ for bimorphic recursion with
no instantiation.

\begin{Def}\rm
The system $\BRNI$ is defined as the type system obtained from
the system $\BR$ by replacing the rule $(rec)$ by the rule $(recni)$:
\[
\infer[(recni)]{U \prove \Rec\{x=e\}:u}{
	U,x:s_1(u) \prove e:u
}
\quad (\Dom(s_1) \subseteq \FTV(u)-\FTV(U))
\]
\end{Def}

This system is an extension of monomorphic recursions where
every recursive call has the same type that is an instantiation of
the type of the function definition.
Since every recursive call has the same type as the type of the function definition
in monomorphic recursion, the system $\BRNI$ can type more expressions
than monomorphic recursion.
For example, the function $\DBtwo$ in Example \ref{eg:1} can be typed
with $\prove \DBtwo:(\beta\List \to \alpha) \to \beta\List \to \alpha$
in this system.

The difference between $(rec)$ and $(recni)$ is that
$(rec)$ has $s_2$ but $(recni)$ does not have $s_2$.
By $(recni)$, the type of a recursively defined function is always
its general type. 
For this reason,
The system types less expressions than our system $\BR$.
For example, the function $\DB$ in Example \ref{eg:1}
cannot be typed because we have to instantiate $\alpha$ by $(\beta\List)$
in the type of $\DBtwo$ in order to type $\DB$.
For the same reason, the system $\BRNI$ does not have the instantiation
property described by Proposition \ref{prop:instantiate}.

We define
\[
(e_1,e_2) = \Pair e_1 e_2,\\
e.1 = \Fst e,\\
e.2 = \Snd e,\\
K = \lambda xy.x,\\
(e_1 \doteq e_2) = \lambda y.(ye_1,ye_2) \qquad (y \not\in \FTV(e_1e_2))
\]

We suppose variables $z_1,z_2,\ldots$ are chosen for type variables
$\alpha_1,\alpha_2,\ldots$.
$\Tilde M$ is defined by $\Tilde \alpha_i = z_i$ and
$\Tilde {M_1 \times M_2} = (\Tilde M_1,\Tilde M_2)$.

Note that when $e_1 \doteq e_2$ is typable, the expressions $e_1$ and $e_2$ have
the same type.
The principal type of $\Tilde M$ is $M$.
When $\Tilde M \doteq \Tilde N$ is typable, we can unify $M$ and $N$.

\begin{Lemma}\label{lemma:1}
(1) $\vec z:\vec u \prove \Tilde M:s(M)$ where $s(\alpha_i)=u_i$.

(2) If $U \prove e:u$, $U \prove e:v$, and $e$ is defined
by $e::=x|\lambda x.e|ee|(e,e)|e.1|e.2$,
then $u=v$.
\end{Lemma}

{\em Proof.}
(1) By induction on $M$.

(2) By induction on $e$.
$\Box$

\begin{Lemma}\label{lemma:wprni}
Let $\vec\alpha = \FTV(M_1,M_2,M_3,M_4)$ and $\vec z = \Tilde{\vec\alpha}$.
Let
\[
e_1 = \Rec\{ f=\lambda \vec z.K(\Tilde M_1,\Tilde M_2)
	(\lambda \vec y.(f\vec y.1 \doteq \Tilde N_1)) \}, \\
e_2 = \Rec\{ f=\lambda \vec z.K(\Tilde M_1,\Tilde M_2)
	(\lambda \vec y.(f\vec y.2 \doteq \Tilde N_2)) \},
\]
where $\vec y$ are fresh variables of the same length as $\vec z$.
The judgment $\prove e_1 \doteq e_2:u$ is provable in $\BRNI$ for some $u$
if and only if
the semiunification problem $\{ M_1 \le N_1, M_2 \le N_2 \}$ has a semiunifier.
\end{Lemma}

We explain proof ideas.
Since $K$ is the constant function combinator,
both $e_1$ and $e_2$ are equal to $\lambda\vec z.(\Tilde M_1, \Tilde M_2)$.
By $e_1 \doteq e_2$, the expressions $e_1$ and $e_2$ have the same type.
Since $f$ is a recursive call, the type of $f$ in the body of the recursive definition in $e_1$ is
some instantiation of the type of $e_1$.
Hence the type of $f\vec y.1$ is some instantiation of the type of $\Tilde M_1$.
Since $f\vec y.1 \doteq \Tilde N_1$, the expressions $f\vec y.1$ and $\Tilde N_1$ is the same type, and therefore
the type of $\Tilde N_1$ is some instantiation of the type of $\Tilde M_1$.
For a similar reason, 
the type of $\Tilde N_2$ is some instantiation of the type of $\Tilde M_2$.

{\em Proof.}
Let 
\[
e_3 = K(\Tilde M_1,\Tilde M_2)
	(\lambda \vec y.(f\vec y.1 \doteq \Tilde N_1)), \\
e_4 = K(\Tilde M_1,\Tilde M_2)
	(\lambda \vec y.(f\vec y.2 \doteq \Tilde N_2)).
\]

From the left-hand side to the right-hand side.
We suppose $\prove e_1 \doteq e_2:u_0$.
Then we have $\prove e_1:u$ and $\prove e_2:u$ for some $u$.

We have $f:s_1(u) \prove \lambda \vec z.e_3:u$ for some $s_1$.
Hence $f:s_1(u),\vec z:\vec u \prove e_3:u'$ and $u=\vec u\imp u'$
for some $u'$ and some $\vec u$.
Let $s(\alpha_i)=u_i$.
Since $\vec z:\vec u \prove (\Tilde M_1,\Tilde M_2):s(M_1 \times M_2)$ by 
Lemma \ref{lemma:1} (1),
we have $u'=s(M_1 \times M_2)$ by Lemma \ref{lemma:1} (2).
Hence $u=\vec u \imp s(M_1 \times M_2)$.
Therefore $f:s_1(u),\vec y:s_1(\vec u) \prove f\vec y.1:s_1s(M_1)$.
Since $\vec z:\vec u \prove \Tilde N_1:s(N_1)$ by Lemma \ref{lemma:1} (1),
we have $s_1s(M_1) = s(N_1)$ by Lemma \ref{lemma:1} (2).

Similarly we have $s_2s(M_2) = s(N_2)$ for some $s_2$.

Hence the semiunification problem $\{ M_1 \le N_1, M_2 \le N_2 \}$ has a semiunifier $s$.

From the right-hand side to the left-hand side.
We suppose $s_1(s(M_1)) = s(N_1)$ and $s_2(s(M_2)) = s(N_2)$.
Let $\vec u$ be $s(\vec \alpha)$, $U$ be $\vec z:\vec u$, and
$u$ be $\vec u \imp s(M_1 \times M_2)$.

By Lemma \ref{lemma:1} (1), we have 
$U \prove \Tilde M_1:s(M_1)$,
$U \prove \Tilde M_2:s(M_2)$,
$U \prove \Tilde N_1:s(N_1)$, and
$U \prove \Tilde N_2:s(N_2)$.

We have $f:s_1(u),\vec y:s_1(\vec u) 
\prove f\vec y.1:s_1(s(M_1))$.
Hence $U, f:s_1(u),\vec y:s_1(\vec u) 
\prove f\vec y.1 \doteq \Tilde N_1:v$ for some $v$.
Hence $U, f:s_1(u) 
\prove \lambda\vec y.(f\vec y.1 \doteq \Tilde N_1):s_1(\vec u) \imp v$.
Combining it with $U \prove (\Tilde M_1, \Tilde M_2):s(M_1\times M_2)$, we
have $U, f:s_1(u) \prove e_3:s(M_1\times M_2)$.
Hence $f:s_1(u) \prove \lambda \vec z.e_3:u$.
By the $(rec)$ rule, we have $\prove \Rec\{f=\lambda\vec z.e_3\}:u$.

Similarly we have $\prove \Rec\{f=\lambda\vec z.e_4\}:u$.
Hence we have $\prove e_1 \doteq e_2:u_0$ for some $u_0$.
$\Box$

\begin{Th}
The typability in $\BRNI$ is undecidable.
\end{Th}

{\em Proof.}
If it were decidable, by Lemma \ref{lemma:wprni}, there would be
an algorithm solving semiunification problems of the form 
$\{ M_1 \le N_1, M_2 \le N_2 \}$.
Since semiunification problems of the form 
$\{ M_1 \le N_1, M_2 \le N_2 \}$ are undecidable by Theorem \ref{th:semiunification},
the typability in $\BRNI$ is undecidable.
$\Box$

The difference between $\BR$ and $\BRNI$ comes from the instantiation property. Since the $(recni)$ rule does not have $s_2$, the system $\BRNI$ does not have
the instantiation property like Proposition \ref{prop:instantiate}.
So we cannot use the same idea for $\BRNI$ since
we cannot replace a uppermost $(recni)$ rule by
\[
\infer*[s(\pi_2)]{}{
\infer[(axiom)]{s(U) \prove \Rec\{x=e\}:s_1(v)}{}
}
\]
for some $s_1$. It is because
$s(U) \prove \Rec\{x=e\}:s_1(v)$ may not be provable for some $s_1$,
even if $s(U) \prove \Rec\{x=e\}:v$ is provable.

\section{Concluding Remarks}

We have proposed the type system $\BR$ with bimorphic recursion.
Bimorphic recursion is restricted polymorphic recursion such that
each recursive call in the body of the function
definition has the same type, and recursive definitions can be nested.
We have proved that this type system has principal types and
decidable type inference.
We have also shown that 
the extension of bimorphic recursion with
the let polymorphism also has principal types and decidable type inference.

Trying to show the decidability of the abstract interpretation given in 
\cite{Gori03} will be a future work. 
We have shown that the type inference of bimorphic recursion is decidable,
and our bimorphic recursion is inspired by \cite{CominiDamiani08}.
By clarifying the relationship among 
the abstract interpretation, the type system
in \cite{CominiDamiani08}, and our bimorphic recursion,
we could show the decidability of the abstract interpretation.

Characterizing a class of semi-unification problems that correspond
to the type inference for our bimorphic recursion will be another
future work.
We can expect the class will be
larger than semi-unification problems with a single inequation. 
The computational complexity of the class would be another future work.

\section*{Acknowledgments}
We would like to thank Prof. Fritz Henglein, Prof. Marco Comini,
Prof. Stefano Berardi, and
Prof. Kazushige Terui for discussions and comments.
We would also like to thank the anonymous referees for valuable comments.


{

\frenchspacing  

}


\begin{thebibliography}{9}

\providecommand{\urlalt}[2]{\href{#1}{#2}}
\providecommand{\doi}[1]{doi:\urlalt{http://dx.doi.org/#1}{#1}} 

\bibitem{CominiDamiani08}
	M. Comini, F. Damiani \& S. Vrech (2008): 
	{\em On Polymorphic Recursion, Type Systems, and Abstract Interpretation}.
	{\em Proceedings of SAS 2008, LNCS} 5079, pp. 144--158,
	\doi{10.1007/978-3-540-69166-2\_10}.

\bibitem{Cop80}
M.~Coppo (1980):
 {\em An extended polymorphic type system}.
 {\em Proceedings of MFCS'80, LNCS} 88, pp. 194--204.

\bibitem{Cous:POPL-1997}
	P. Cousot (1997): {\em Types as abstract interpretation}.
	{\em Proceeding of POPL 97}, pp. 316--331,
	\doi{10.1145/263699.263744}.

\bibitem{DamMil82}
	L. Damas \& R. Milner (1982):
	{\em Principal type schemes for functional programs}.
	{\em Proceedings of POPL 82}, pp. 207--212,
	\doi{10.1145/582153.582176}.

\bibitem{Damiani:FI-07}
F.~Damiani (2007):
{\em Rank 2 intersection for recursive definitions}.
 {\em Fundamenta Informaticae} 77(4), pp. 451--488.

\bibitem{GoriLevi:VMCAI-2002}
R.~Gori \& G.~Levi (2002):
 {\em An experiment in type inference and verification by abstract
  interpretation}.
 {\em Proceedings of  VMCAI'02, LNCS} 2294, pp. 225--239.

\bibitem{Gori03}
	R. Gori \& G. Levi (2003):
	{\em Properties of a type abstract interpreter}.
	{\em Proceedings of VMCAI'03,
	LNCS} 2575, pp. 132--145,
	\doi{10.1007/3-540-36384-X\_13}.

\bibitem{Henglein89}
	F. Henglein (1989):
	{\em Polymorphic Type Inference and Semi-Unification}.
	Ph.D. thesis, the state university of New Jersey.

\bibitem{Henglein93}
F.~Henglein (1993):
 {\em Type inference with polymorphic recursion}.
 {\em ACM TOPLAS} 15(2), pp. 253--289.

\bibitem{Hin97}
R.~Hindley (1997):
 {\em {Basic Simple Type Theory}}.
  Cambridge University Press.

\bibitem{Jim96}
T.~Jim (1996):
 {\em What are principal typings and what are they good for?}
 {\em Proceedings of POPL'96}, pp. 42--53.

\bibitem{Kfo+Per:LICS-1999}
A.~J. Kfoury \& S.~M. Pericas-Geertsen (1999):
 {\em Type inference for recursive definitions}.
 {\em Proceedings of LICS'99}, pp. 119--128,
 \doi{10.1109/LICS.1999.782600}.

\bibitem{Kfoury93}
	A.J. Kfoury, J. Tiuryn, \& P. Urzyczyn (1993):
	{\em Type Reconstruction in the Presence of Polymorphic Recursion}.
	{\em ACM TOPLAS} 15 (2),
	pp. 290--311.

\bibitem{Mee83}
L.~Meertens (1983):
 {\em Incremental polymorphic type checking in {B}}.
 {\em Proceedings of POPL'83}, pp. 265--275.

\bibitem{Monsuez92}
B.~Monsuez (1992):
 {\em Polymorphic typing by abstract interpretation}.
 {\em Theoretical Computer Science} 652, pp. 217--228.

\bibitem{Monsuez:SAS-1993}
B.~Monsuez (1993):
 {\em Polymorphic types and widening operators}.
 {\em Proceedings of SAS'93, LNCS} 724, pp. 224--281.

\bibitem{Myc84}
A.~Mycroft (1984):
 {\em Polymorphic Type Schemes and Recursive Definitions}.
 {\em LNCS} 167, pp. 217--228.

\bibitem{Rittri95}
	M. Rittri (1995):
	{\em Dimension inference under polymorphic recursion}.
	{\em Proceedings of FPCA '95}, pp. 147--159.
	
\bibitem{Ter+Aik:LICS-2006}
T.~Terauchi \& A.~Aiken (2006):
 {\em On typability for polymorphic recursive rank-2 intersection types}.
 {\em Proceedings LICS'06}, pp. 111--122,
 \doi{10.1109/LICS.2006.41}.

\end{thebibliography}
\end{document}